# Self-dual quaternionic lumps in octonionic space-time

Graeme D. Robertson[1]
*Department of Applied Mathematics and Theoretical Physics, University of Cambridge, Cambridge, UK*

11 February 2003

The theory of self-dual bosonic lumps immersed in the Cayley-calibrated space of octonions has a large class of exact finite quaternionic power series solutions.

**1. Bosonic *p*-brane dynamics**

The general theory of *p*-dimensionally extended supersymmetric objects and their classification [1] has guided the search for M-theory. Super *p*-branes originate from a generalization of the Green-Schwarz-Witten superstring action which itself came from an equivalent of the generalised Einstein-Dirac-Nambu-Goto action. This action involves a Lagrangian which is proportional to the world (*d*=*p*+1)-volume swept out by the extended object in a *D*-dimensional background.

$$S[X(\xi)] = M \int \sqrt{g} \, d^d \xi \tag{1.1}$$

where

$$g(X) = \det g_{ab}$$

and $g_{ab}$ is the induced metric in the p-brane's world subspace

$$g_{ab}(X) \equiv X^{\mu}_{,a} X_{\mu,b}$$

using the notation

$$X^{\mu}_{,a}(\xi) \equiv \frac{\partial X^{\mu}}{\partial \xi^a}$$

*a,b*=1,2,..,*d* and *μ*=1,2,..,*D*.

Minimizing the world *d*-volume swept out by the *p*-brane in a flat Euclidean background yields the Euler-Lagrange equation of motion

$$\frac{1}{\sqrt{g}} \partial_a (\sqrt{g} \, g^{ab} X^{\mu}_{,b}) = 0$$

which is the Laplacian of $X^{\mu}$ in general coordinates. By analogy with Newton's law $d_t p = F$, the generalised momentum is defined as

$$P^{a\mu} \equiv M \sqrt{g} \, g^{ab} X^{\mu}_{,b} \tag{1.2}$$

where $M$ is a generalised invariant mass density with dimensions [M L$^{1-p}$ T$^{-1}$]. The equation of motion can then be written

---
[1] Visitor

$$\partial_a P^{a\mu} = 0 \tag{1.3}$$

The appropriate identities, from world volume reparametrisation invariance, can be written

$$P^a_\mu X^\mu_{,b} - M\sqrt{g}\,\delta^a_b = 0 \tag{1.4}$$

which corresponds to the vanishing of the Hamiltonian, $p.\mathrm{d}_t x - L = H$, and

$$P^a_\mu P^{b\mu} = M^2\, g\, g^{ab} \tag{1.5}$$

which is the extended object generalisation of $p^2 = m^2$ for relativistic particles.

## 2. Self-dual lumps

Not all classes of $p$-brane immersions admit a dual formulation. Examples which do are string ($p=1$) in 4 dimensions with an almost complex structure, ($d=2;D=4$)-brane [2], membrane ($p=2$) in 4 dimensions, (3;4)-brane [3], membrane in 7 (associative calibration) dimensions, (3;7)-brane [4], and lump ($p=3$) in 8 (Cayley calibrated) dimensions [5], as well as ($d;d$)-branes [6].

Consider the case of (4;8)-brane. Self-duality, analogous to self-duality in the bosonic sector of Yang-Mills gauge fields, can be formulated as

$$F^{\mu\nu}_{ab} = \frac{1}{4}\varepsilon_{ab}{}^{cd}\, T^{\mu\nu}{}_{\rho\sigma}\, F^{\rho\sigma}_{cd} \tag{2.1}$$

where

$$F^{\mu\nu}_{ab} \equiv X^\mu_{,[a}\, X^\nu_{,b]}$$

$\varepsilon^{abcd}$ is the 4-dimensional permutation tensor and $T_{\mu\nu\rho\sigma}$ is a completely anti-symmetric tensor [7] that defines a basis for octonionic multiplication.

Contracting (2.1) with $X_\nu{}^{,b}$ gives

$$X^\mu_{,a} = \frac{1}{3!}\varepsilon_a{}^{bcd}\, T^\mu{}_{\nu\rho\sigma}\, X^\nu_{,b}\, X^\rho_{,c}\, X^\sigma_{,d} \tag{2.2}$$

Combining this with (1.2) gives the self-dual world-volume momentum current density

$$P^\mu_{,a} = \frac{1}{3!} M\, e_a{}^{bcd}\, T^\mu{}_{\nu\rho\sigma}\, X^\nu_{,b}\, X^\rho_{,c}\, X^\sigma_{,d} \tag{2.3}$$

If this expression for $P^{a\mu}$ is substituted into the equation of motion (1.3) then the equation is satisfied automatically, as can be seen from the symmetry of $\partial_a X^\nu{}_{,b}$ versus the anti-symmetry in indices $a$ and $b$ of the permutation symbol $e^{abcd}$.

Contracting (2.2) with $X_\mu{}^{,a}$ yields

$$\sqrt{g} = T_{\mu\nu\rho\sigma}\, X^\mu{}_{,1}\, X^\nu{}_{,2}\, X^\rho{}_{,3}\, X^\sigma{}_{,4} \qquad (2.4)$$

which is the condition necessary for (2.3) to satisfy (1.4). Substituting expression (2.3) twice in $P^a{}_\mu P^{b\mu}$ can be simplified to $M^2 g\, g^{ab}$, showing that (1.5) is automatically satisfied by (2.3). Thus all the constraints and equation of motion of lumps in 8 octonionic dimensions are solved automatically by the self-dual construction of $P^{a\mu}$ in (2.3).

## 3. Quaternionic solutions

Solving (2.4) will ensure that (2.3) satisfies all the constraints, as well as the equation of motion, of (4;8)-brane. In order to find explicit solutions of (2.4), it is necessary to choose one particular basis for quaternionic and octonionic multiplication. There are only two possibilities for quaternions; often understood as left-handed and right-handed coordinate systems. Here, as usual, the right-handed system is chosen. It is fixed by specifying that $e_{123} = +1$, or $i.j=k$, and is simply written as 1 2 3.

For octonions there are 480 admissible bases, 240 clockwise and 240 anti-clockwise. A straight-forward choice is the basis characterised by the seven triples, (1 2 3)(2 4 6)(3 6 5)(4 5 1)(5 7 2)(6 1 7)(7 3 4) which determine $T_{\mu\nu\rho\sigma}$ uniquely [5].

Heretofore, the only known non-trivial solutions to the self-dual (4;8)-brane involve 4 arbitrary complex functions of pairs of lump coordinates [5]. The first quaternionic example of a solution to (2.4), and hence to all the requirements for lumps in 8 octonionic dimensions, including (2.1), is the trivial case of $\xi$, considered as a quaternion of lump coordinates. Take $\xi$ to be the first quaternionic element of the two (quaternion-valued) component octonion, $X$.

$$X = (\xi_1, \xi_2, \xi_3, \xi_4, 0, 0, 0, 0) \qquad (3.1)$$

written as

$$X = (\xi\,; 0)$$

This choice of $X$ leads to the constant world volume velocity matrix

$$X^\mu{}_{,a} = \begin{pmatrix} 1 & 0 & 0 & 0 & 0 & 0 & 0 & 0 \\ 0 & 1 & 0 & 0 & 0 & 0 & 0 & 0 \\ 0 & 0 & 1 & 0 & 0 & 0 & 0 & 0 \\ 0 & 0 & 0 & 1 & 0 & 0 & 0 & 0 \end{pmatrix} \qquad (3.2)$$

From (3.2) it follows that the induced metric is the unit matrix and the world volume $\sqrt{g} = 1$.

The world volume momentum, $P^\mu{}_a$, is calculated from (2.3). Taking units in which $M=1$, and using the basis for $T_{\mu\nu\rho\sigma}$ as specified above, this matrix is the same as that in (3.2) above, and thus $P^\mu{}_a$ clearly satisfies (1.4) and (1.5). These identities could be verified simultaneously by the condition that (2.4) is satisfied by the $X^\mu$ of (3.1).

Consider now the case where $\xi$ is replaced with the quaternionic square of $\xi$,

$$\xi^2 = (\xi_1^2 - \xi_2^2 - \xi_3^2 - \xi_4^2, 2\xi_1\xi_2, 2\xi_1\xi_3, 2\xi_1\xi_4)$$

Then

$$X = (\xi^2 ; 0)$$

and

$$X^{\mu}{}_{,a} = 2 \times \begin{pmatrix} \xi_1 & \xi_2 & \xi_3 & \xi_4 & 0 & 0 & 0 & 0 \\ -\xi_2 & \xi_1 & 0 & 0 & 0 & 0 & 0 & 0 \\ -\xi_3 & 0 & \xi_1 & 0 & 0 & 0 & 0 & 0 \\ -\xi_4 & 0 & 0 & \xi_1 & 0 & 0 & 0 & 0 \end{pmatrix}$$

(3.3)

Using this in (2.3) yields

$$P^{a\mu} = 8\xi_1 \times \begin{pmatrix} \xi_1^2 & \xi_1\xi_2 & \xi_1\xi_3 & \xi_1\xi_4 & 0 & 0 & 0 & 0 \\ -\xi_1\xi_2 & (\xi_1^2+\xi_3^2+\xi_4^2) & 0 & 0 & 0 & 0 & 0 & 0 \\ -\xi_1\xi_3 & 0 & (\xi_1^2+\xi_2^2+\xi_4^2) & 0 & 0 & 0 & 0 & 0 \\ -\xi_1\xi_4 & 0 & 0 & (\xi_1^2+\xi_2^2+\xi_3^2) & 0 & 0 & 0 & 0 \end{pmatrix}$$

(3.4)

Substituting $X^{\mu}{}_{,a}$ into (2.4) is found to satisfy the equation, with

$$\sqrt{g} = T_{\mu\nu\rho\sigma} X^{\mu}{}_{,1} X^{\nu}{}_{,2} X^{\rho}{}_{,3} X^{\sigma}{}_{,4} = 16\xi_1^2(\xi_1^2 + \xi_2^2 + \xi_3^2 + \xi_4^2)$$

which implies that all constraints, as well as the equation of motion and the self-dual equation (2.1), are satisfied.

All solutions of the form $X = (\xi^n ; 0)$ for n=1 to 8 have been validated by analytic computation, which suggests that $X = (\xi^n ; 0)$ is a solution for all positive integer n.

More interesting solutions would have non-zero elements in all components. $T_{\mu\nu\rho\sigma}$ is a tensor which is invariant under SO(8). Transforming $X^{\mu}$ under SO(8) should preserve the solution while altering its component form. However, with analytic computation it is possible to pursue a more empirical approach. Following the observation in [2] that interesting solutions can be generated by swapping the order of real and imaginary parts of two complex functions *u* and *v* when forming a 4 component object (*Im u*, *Re u*, *Re v*, *Im v*), various permutations of quaternion components can be plugged into X and tested to see if they solve (2.4).

It is found that, in the octonion basis, (1 2 3)(2 4 6)(3 6 5)(4 5 1)(5 7 2)(6 1 7)(7 3 4),

$$X = (\xi_1, \xi_2, \xi_3, \xi_4, \xi_4, \xi_1, \xi_2, \xi_3)$$

solves (2.4). This particular rearrangement of terms (4 1 2 3) in the second half of $X$ will be indicated by a transpose, thus:

$$X = (\xi ; \xi^T)$$

(3.5)

In this case, $g_{ab}$ is twice the unit matrix and $\sqrt{g}$ is 4, as is $T_{\mu\nu\rho\sigma}X^\mu{}_{,1} X^\mu{}_{,2} X^\mu{}_{,3} X^\mu{}_{,4}$. Similarly, $(\xi^2 ; (\xi^2)^T)$ solves (2.4). Indeed, all octonions of the form $X = (\xi^n ; (\xi^n)^T)$, for $n=1$ to 8, have been shown to solve exactly all the equations of (4;8) lumps.

## 4. General quaternionic solution

Solutions found so far may be superposed to form new solutions.

$$X = (\sum_{n=1}^{3} a_n \xi^n ; 0)$$

(4.1)

where $a_n$ are arbitrary real constants, has been shown to solve all the equations, in particular (2.4).

Further, superposed octonions of type (3.5) are also solutions. Octonionic functions of the form

$$X = (\sum_{n=1}^{3} a_n \xi^n ; (\sum_{n=1}^{3} a_n \xi^n)^T)$$

(4.2)

where $a_n$ are real coefficients, solve all lump equations.

Note that the power series coefficients, $a_n$, must be the same in both halves of $X$. If arbitrary coefficients $b_n$ are inserted in place of $a_n$ in the second half of $X$ then (2.4) is not satisfied. For example, with $X$ defined as

$$X = (\sum_{n=1}^{2} a_n \xi^n ; (\sum_{n=1}^{2} b_n \xi^n)^T)$$

the difference is given by

$$\sqrt{g} - T_{\mu\nu\rho\sigma} X^\mu{}_{,1} X^\nu{}_{,2} X^\rho{}_{,3} X^\sigma{}_{,4} = 4(a_1 b_2 - a_2 b_1)^2 (\xi_2 + \xi_4)^2$$

This result would seem to prohibit the possibility of finding knotted quaternionic instantons as sought in [8].

All functions of the form (4.2), where $n = 1,2,...,\infty$ (i.e. any positive integer), are conjectured to solve the lump equations. A potentially interesting subset of these can be written

$$X = (a\, e^\xi ; (a\, e^\xi)^T)$$

Solution (4.2) can be generalised in one further way. Each half of $X$ can be multiplied by a real constant coefficient. The general solution

$$X = (c_1 \sum_{n=1}^{\infty} a_n \xi^n ; c_2 (\sum_{n=1}^{\infty} a_n \xi^n)^T)$$

(4.3)

has been verified up to $n=3$ with arbitrary real coefficients $c_1$ and $c_2$.

## 5. Comments

The motion of a material particle under the action of gravitation can be derived from Einstein's action integral

$$S = m \int \sqrt{g_{\mu\nu} \, dX^\mu \, dX^\nu}$$

(5.1)

As a direct result of the square root in (5.1), the energy of a particle combines with its momentum to form the well-known invariant

$$m = \pm\sqrt{E^2 - p^2}$$

It is this energy-momentum scalar which Dirac managed to linearise by introducing a matrix representation of the square root. The elementary quantum solutions have spin ½ (an SU(2) symmetry) and are called fermions. Supersymmetry is introduced in order to unify fermions and bosons. If this square root can be removed then the need for a Dirac square root, and ultimately the need for supersymmetry, would be obviated.

As suggested in [5], equation (2.4) can be used to replace $\sqrt{g}$ in (1.1) giving

$$S = M \int T_{\mu\nu\rho\sigma} \, X^\mu_{,1} \, X^\nu_{,2} \, X^\rho_{,3} \, X^\sigma_{,4} \, d^4\xi$$

which can be rewritten as

$$S = \frac{1}{4!} M \int T_{\mu\nu\rho\sigma} \, dX^\mu \, dX^\nu \, dX^\rho \, dX^\sigma$$

(5.2)

in analogy with (5.1), but without the explicit square root.

Restricting consideration to bosonic lumps from the outset is therefore not necessarily a shortcoming of this final theory of lumps. Supersymmetric lumps (slumps) may not be required.

The theory described by (5.2) is based on a principle of maximization of associativity of quaternions of lump coordinates in octonionic space rather than the minimization of the swept-out lump 4-volume.

It remains to be seen whether it is possible to find suitable representations for the elementary fields of the standard $SU(3)_C \otimes SU(2)_L \otimes U(1)_Y$ model in the rich group structure of the non-commutative quaternion field, whose algebra is closely connected to the Lie algebra of

$$SO(3) \cong SU(2) \supset U(1)$$

and the non-associative octonion field, whose algebra is closely connected to the Lie algebra of

$$SO(8) \supset SO(7) \supset G_2 \supset SU(3) \supset SU(2) \otimes U(1)$$

This leaves plenty of scope for Kaluza-Klein-type compactifications.

One last comment might be of interest. The Dirac square root introduces negative energy solutions, which are interpreted as antimatter. In the above theory, the general solution (4.3) has a distinct correspondence between components $\sum a_n \, \xi^n$ of $X$ and components $(\sum a_n \, \xi^n)^T$ of $X$. This tight parallelism might be interpretable as the matter - anti-matter symmetry. There is also a gauge freedom in the $c$ constants, which will lead to a conserved quantity. Elementary mass?

**Acknowledgement**

It is a great pleasure to thank all my teachers, especially James Rainy Brown, Brian Thompson, Mary Hesse and Ed Corrigan for their clear and precise explanations of science. It is also a pleasure to thank DAMTP for their kind hospitality.

**References**

[1] A.Achúcarro, J.M.Evans, P.K.Townsend and D.L.Wiltshire, *Super p-branes*, Phys. Lett. **198B** (1987) 441
[2] G.D.Robertson, *Torus knots are rigid string instantons*, Phys. Lett. **226B** (1989) 244
[3] B.Biran, E.G.Floratos and G.K.Savvidy, *The self-dual closed bosonic membranes*, Phys. Lett. **198B** (1987) 329
[4] M.Grabowski and C-H.Tze, *Generalized self-dual bosonic membranes, vector cross products and analyticity in higher dimensions*, Phys. Lett. **224B** (1989) 329
[5] G.D.Robertson, *Self-dual lumps and octonions*, Phys. Lett. **249B** (1990) 216
[6] R.P.Zaikvo, *Self-duality in the theory of the bosonic p-branes*, Phys.Lett. **211B** (1988) 281
[7] E.F.Corrigan, C.Devchand, D.B.Fairlie and J.Nuyts, *First-order equations for gauge fields in spaces of dimension greater than four*, Nucl. Phys. **B 214** (1983) 452
[8] G.D.Robertson, *Self-dual quaternionic lumps in octonionic space-time*, DTP-89/39A